\documentclass[aps,prb,preprint,showpacs,groupedaddress]{revtex4}
\usepackage{amssymb}

\usepackage{graphicx}

\begin{document}

\title[]{Magnetic Field Effect and Dielectric Anomalies at the Spin Reorientation Phase Transition of GdFe$_3$(BO$_3$)$_4$}

\author{F. Yen$^1$, B. Lorenz$^1$, Y. Y. Sun$^1$, and C. W. Chu$^{1,2,3}$, L. N. Bezmaternykh$^4$, A. N. Vasiliev$^5$}

\affiliation{$^1$ TCSUH and Department of Physics, University of
Houston, Houston, Texas 77204-5002, USA}

\affiliation{$^2$ Lawrence Berkeley National Laboratory, 1 Cyclotron
Road, Berkeley, California 94720, USA}

\affiliation{$^3$ Hong Kong University of Science and Technology,
Hong Kong, China}

\affiliation{$^4$ Institute of Physics, Siberian Division, Russian
Academy of Sciences, Krasnoyarsk, 660036 Russia}

\affiliation{$^5$ Faculty of Physics, Moscow State University,
Moscow, 119992 Russia}

\begin{abstract}
GdFe$_3$(BO$_3$)$_4$ exhibits a structural phase transition at 156
K, antiferromagnetic order of the Fe$^{3+}$ moments at 36 K followed
by a spin reorientation phase transition at 9 K. The reorientation
phase transition is studied through dielectric, magnetic and heat
capacity measurements under the application of external magnetic
fields of up to 25 kOe. The dielectric constant indicates the
existence of two distinct anomalies at T$_{SR}$ = 9 K that separate
in temperature under external magnetic fields. The spin rotation
phase transition is proven to be of the first order nature through
the magnetic analogue of the Clausius-Clapeyron equation.
Magneto-dielectric effect of up to 1 percent is observed at 8 K and
7 kOe. Uniaxial magnetocaloric effect along the c-axis is observed
below the spin reorientation phase transition of 9 K.
\end{abstract}



\maketitle

GdFe$_3$(BO$_3$)$_4$ belongs to the trigonal system with space group
R32. It is similar to huntite CaMg$_3$(CO$_3$)$_4$, a trigonal
trapezohedral structure that is one of the five trigonal types.
Borate crystals of this category are appealing because of their
possible applications as single crystal minilasers due to their good
luminescent and nonlinear optical properties. In particular for
GdFe$_3$(BO$_3$)$_4$ single crystals, recent studies have focused on
the better understanding of its optical properties and at the same
time on its magnetic properties through phase matching of absorption
and second harmonic generation spectra.$\cite{1, 2}$ The magnetic
properties of the rare-earth iron borates are also of fundamental
interest because of the existence and mutual interference of two
magnetic subsystems (Fe and Gd). It is of fundamental importance the
understanding of the magnetic orders of the gadolinium and iron
sublattices and the coupling between the iron spins and the
gadolinium moments which contributes to the crystal's rich magnetic
properties.

The crystal structure of huntite has been analyzed and described
elsewhere$\cite{3, 4, 5}$, it consists of GdO$_6$ bipyramids and
FeO$_6$ octahedrons. The octahedrons form 3-fold helicoidal chains
along the c-axis. The bipyramids are located between nearly three
equal distant octahedrons. Rare-earth iron borates,
RFe$_3$(BO$_3$)$_4$ (R=Eu to Ho and Y), undergo a structural
transition at higher temperatures and an antiferromagnetic (AFM)
transition involving the Fe spins at T$_N$$\approx$35 K.$\cite{6}$
For R=Gd a weakly first order structural phase transition at
T$_1$=156 K changes the structural symmetry from R32 to P3$_1$2$_1$.
$\cite{5}$ A second order phase transition at T$_N$=36 K results in
the AFM ordering of the Fe$^{3+}$ magnetic moments aligned in the
basal plane. At T$_{SR}$=9 K, another sharp phase transition occurs
characterized by a spin reorientation of the Fe$^{3+}$ magnetic
moments by 90 degrees from the basal plane to the c-axis.

The coupling between the Gd moments and the Fe spins at low
temperatures is strong in GdFe$_3$(BO$_3$)$_4$ and it was speculated
that the reorientation of the AFM iron spin system is triggered by
the anisotropy of the Gd moments aligned with the c-axis.$\cite{6}$
This exchange interaction is indirect and involves other ions such
as oxygen in the structure. Below T$_N$, with decreasing
temperature, the magnetic order grows and that causes strain via the
spin-lattice interaction. At the spin reorientation phase transition
the magnetic order experiences a sudden change which is reflected in
distinct anomalies of the dielectric constant, $\epsilon$, as
recently discussed in several rare-earth manganites.$\cite{7, 8}$
The dielectric constant can be measured with extraordinary precision
and it monitors subtle changes of the magnetic system. We have
therefore searched for dielectric anomalies at the magnetic phase
transitions in GdFe$_3$(BO$_3$)$_4$ and its correlation with
magnetic and heat capacity data. Below T$_N$, close to T$_{SR}$, we
found two distinct anomalies of $\epsilon$ that separate in
temperature with applied magnetic fields. A sharp drop of $\epsilon$
occurs at the reorientation of the Fe spins whereas a maximum of
$\epsilon$(T) above T$_{SR}$ indicates the onset of magnetic field
induced ferroelectric order. The spin reorientation phase transition
is proven to be of the first order nature through the magnetic
Clausius-Clapeyron equation. Lastly, a phase diagram is proposed for
magnetic fields parallel or perpendicular to the crystalline c-axis.

The single crystal GdFe$_3$(BO$_3$)$_4$ was grown as described
elsewhere.$\cite{9}$ The transparent crystal (shining green in
color) was analyzed and oriented using a GADDS x-ray diffractometer.
The heat capacity data was acquired via Quantum Design's Physical
Property Measurement System (PPMS) under the application of magnetic
fields of up to 6 kOe both along $a$- and $c$-axis (we use the
hexagonal coordinate system in which $a\perp c$). DC magnetization
data was measured employing Quantum Design's Magnetic Property
Measurement System (MPMS) in magnetic fields parallel to $a$ and
$c$. Smaller portions of the crystal were cut from an original
crystal in order to align the two desired crystallographic
orientations for dielectric investigations. Silver paint was used as
electrodes for the dielectric measurements and the sample was
mounted onto a home-made capacitance probe that was adapted to the
PPMS. The capacitance was measured by the high precision capacitance
bridge AH2500A (Andeen Hagerling) operating at a frequency of 1 kHz.
The temperature and magnetic fields were controlled by the PPMS when
the dielectric constant was measured.

The three known phase transitions are reflected in different
anomalies of the dielectric constant (Fig. 1). At T$_1$=156 K, the
dielectric constant drops significantly since the transition is a
structural one. T$_1$ does not depend on magnetic field. The
magnetic transition into the AFM2 phase at T$_N$ is distinguished by
a change in slope of $\epsilon_c$ and a minimum of $\epsilon_a$ at
36 K. Below T$_N$, $\epsilon_a$ starts increasing with decreasing T
and develops the peak-like feature at T$_{SR}$, the transition into
the AFM1 phase, as shown in Fig. 1. This is a clear signature of
lattice softening and it has to be correlated with the changes in
the magnetic order at and below T$_N$. A similar, but positive slope
change at T$_N$ is observed in $\epsilon_c$. This behavior is
analogous to the magnetic data which will be discussed later. In a
system where there are two magnetic subsystems present with one of
them being a d-metal and the other one an f-metal subsystem, the
f-metal is magnetically polarized upon the ordering of the d-metal
subsystem.$\cite{10}$ In GdFe$_3$(BO$_3$)$_4$  the gadolinium
develops an AFM order along the c-axis through inter planar exchange
interactions mediated by the Gd-O-Fe bonds below T$_N$. The
dielectric anomalies observed at the magnetic phase transitions are
an indication of strong spin-lattice interactions and are attributed
to magneto-elastic effects similar to those observed in some rare
earth manganites$\cite{11}$ and a strong coupling of the dielectric
response with the rare earth magnetic moment was also reported in
these compounds.$\cite{7,8}$ The softness of the lattice (as
expressed by the sharp increase of $\epsilon_a$(T)) is therefore an
indication of the softening of the magnetic order as T approaches
T$_{SR}$ resulting in the Fe-spin reorientation triggered by the Gd
moments and the abrupt decrease of $\epsilon_a$. Thermal hysteresis
was observed in $\epsilon$(T) above T$_{SR}$ at zero magnetic field.
The hysteretic behavior of $\epsilon_a$(T) extends over several
degrees just above T$_{SR}$. Upon cooling in zero field the
dielectric constant below T$_N$ increases continuously for T
$\rightarrow$ T$_{SR}$ and drops rapidly right below T$_{SR}$.
However, upon heating, after the sharp increase at T$_{SR}$ there
appears a distinct maximum of $\epsilon_a$ at T$_M$ about 0.7 K
higher in temperature (Fig. 2). This second anomaly at T$_M$ needs
to be explored in more detail.

The external magnetic field, H, shows an interesting effect on both
dielectric anomalies (T$_{SR}$ and T$_M$). If H is aligned with the
c-axis the spin reorientation transition is suppressed and T$_{SR}$
quickly decreases with H$_c$ in accordance with recent magnetic
data.$\cite{9}$ However, T$_M$ is barely affected by H$_c$ and the
separation of both anomalies increases in magnetic fields as shown
in Fig. 2. In contrast, for magnetic fields parallel to $a$,
T$_{SR}$ remains constant but T$_M$ shifts to higher temperature
(Fig. 3). At the same time the enhancement of $\epsilon_a$(T) is
largely reduced and disappears for fields exceeding 7 kOe. This
qualitative behavior of the dielectric constant in magnetic fields
is closely correlated with the magnetic properties of the coupled
system of the Fe spins and the Gd moments. It is interesting to note
that recent measurements of the magnetic field-induced electric
polarization, P, of GdFe$_3$(BO$_3$)$_4$ have shown a sudden
increase of P in crossing T$_{SR}$ with increasing field.$\cite{21}$
This may indicate the existence of a field-induced ferroelectric
phase (FIP) and it was proposed that higher order magneto-electric
couplings are responsible for the observed phenomena. From our
dielectric measurements we can uniquely identify the boundaries of
this new phase by the sudden drop of $\epsilon_a$(T) at T$_{SR}$ and
the distinct maximum of $\epsilon_a$(T) at the high temperature end
(T$_M$). Therefore, the phase diagram of GdFe$_3$(BO$_3$)$_4$
includes three distinct phases (AFM2, AFM1, and FIP) below the AFM
ordering temperature, T$_N$, of the iron system.

The results of magnetization measurements with fields oriented
parallel to $a$ and $c$ are summarized in Fig. 4. Surprisingly, the
$a$- and $c$-axis DC susceptibilities are equal above T$_N$ within
the experimental resolution, i.e. GdFe$_3$(BO$_3$)$_4$ is
magnetically isotropic at high temperatures although there is
evidence for easy plane anisotropy of the Fe spins and easy axis
anisotropy of the Gd spins.$\cite{6}$ The anisotropies of both
magnetic ions are obviously correlated resulting in a complete
isotropic balancing just like some rare-earth metal ferrite-garnets
and orthoferrites.$\cite{12}$ The high temperature Curie-Weiss
extrapolation yields a Curie-Weiss temperature of $\Theta_{CW}=-34$
K and an effective magnetic moment $\mu$$_{eff}=12.2$ $\mu$$_B$ in
good agreement with other reports$\cite{13}$ and the theoretically
expected value of $\mu$$_{eff}=12.96$ $\mu$$_B$ (Gd$^{3+}$, S=7/2
and Fe$^{3+}$, S=5/2). Deviations from the Curie-Weiss behavior due
to short range AFM correlations are detected below 100 K.

Below T$_N$ the magnetic response becomes anisotropic and depends on
the direction of the probing field. For H $\parallel$ $c$ the DC
susceptibility, $\chi_c$(T), does not exhibit any anomaly at T$_N$
and it is independent of the magnetic field (up to 10 kOe) between
T$_N$ and T$_{SR}$. $\chi_c$ decreases suddenly at T$_{SR}$ as a
result of the spin reorientation aligning the Fe-spins with the
$c$-axis (Fig. 4). The major effect of H$_c$ is the shift of
T$_{SR}$ to lower T and eventually the suppression of the spin
reorientation for fields above 8 kOe. It is obvious that the
magnetic response strongly depends on the angle between the external
field and the Fe spins. This is justifiable since it is
energetically unfavorable to keep the Fe spins aligned with the
external magnetic field. This qualitative behavior is reflected in
the field dependence of the dielectric constant (Fig. 2).
$\epsilon_a$(T) increases below T$_N$ and exhibits a distinct
maximum at T$_M$$\approx$10 K. The sharp drop of $\epsilon_a$(T) at
lower temperature coincides with T$_{SR}$(H) as determined
magnetically. $\epsilon_a$(T) is only weakly dependent on the field
H$_c$ between T$_N$ and T$_M$ as is the magnetic susceptibility,
$\chi_c$. Whereas the value of T$_M$ is not affected by the field
oriented along the $c$-axis T$_{SR}$(H) decreases rapidly reaching
zero at about 8 kOe and separating the two dielectric anomalies in
the magnetic field. The peak of $\epsilon_a$(T) indicating the phase
boundary between the AFM2 phase and the FIP phase exists well above
8 kOe, the critical field that suppresses the AFM1 phase. Details of
the phase diagram will be discussed later.

The magnetic response to an in-plane magnetic field, H$_a$, is
completely different. At low magnetic fields the susceptibility
$\chi_a$ levels off below T$_N$ and is clearly lower than $\chi_c$
(Fig. 4). This can be attributed to the AFM order of the Fe spins
below T$_N$ reducing the susceptibility for H $\parallel$ $a$. Close
to T$_{SR}$, however, $\chi_a$ increases again and its temperature
dependence below T$_{SR}$ appears to be a continuation of the
$c$-axis susceptibility, $\chi_c$ (discussed above). This behavior
agrees well with the observation that the magnetic susceptibility is
little affected by the AFM order if the field is perpendicular to
the Fe spins. At T$_{SR}$ the Fe spins rotate towards the $c$-axis
and align themselves with the magnetic field. $\chi_a$(H) rapidly
increases with H in the temperature range T$_{SR}$$<$T$<$T$_N$ and
it approaches the values of $\chi_c$ above 7 kOe. T$_{SR}$ is
independent of H$_a$, as is also reflected in the dielectric data of
Fig. 3 (the sharp drop of $\epsilon_a$(T) marks the spin
reorientation transition). $\epsilon_a$(T) between T$_{SR}$ and
T$_N$, however, is strongly affected by H$_a$ and its enhancement
below T$_N$ is reduced by the magnetic field. Thereby the maximum of
$\epsilon_a$(T) is shifted to higher temperatures (Fig. 3) reaching
the Neel temperature at a relatively low field of 4 to 5 kOe. This
limits the AFM2 phase and extends the temperature range of the FIP
phase to T$_N$=36 K.

The dielectric anomalies observed in GdFe$_3$(BO$_3$)$_4$ at the
magnetic transitions reveal an interesting correlation between the
different magnetic subsystems (Fe and Gd) and the spin-lattice
coupling leading to the modifications of $\epsilon$(T). To arrive at
a deeper understanding the magnetic correlations have to be
considered in more detail. The phase diagram and the magnetic order
of GdFe$_3$(BO$_3$)$_4$ was recently investigated by AFM resonance
experiments and it was suggested that the Fe spins at high
temperatures are aligned in the $a$-$b$ plane (easy plane
anisotropy) whereas the Gd moment experience a strong easy axis
anisotropy orienting them along the $c$-axis.$\cite{6}$ The AFM
order of the iron moments below T$_N$ couples to the gadolinium
moments resulting in an AFM alignment of the Gd spins ("magnetic
polarization"). With decreasing temperature the Fe-Gd coupling grows
stronger and eventually triggers the reorientation of the Fe spins
resulting in a transition from the non-collinear
(T$_{SR}$$<$T$<$T$_N$) to a collinear (T$<$T$_{SR}$) magnetic order
of both subsystems along the $c$-axis. This phenomenon, related to
the strong single ion anisotropy of rare earth ions, is not uncommon
and it was observed in a number of different compounds involving
rare earth ions and d-elements.$\cite{14,15,16}$ Depending on the
strength of the anisotropy factors the coupling between the
f-moments and d-spins results in the rotation of the d-spins towards
the f-moments (such as in R$_2$CuO$_4$, R=Pr, Nd, Sm,
Eu)$\cite{14,17}$ or vice versa (e.g. in orthorhombic
HoMnO$_3$).$\cite{15}$ In rare cases the f-d exchange can result in
an in-plane rotation of the d-spins as observed in hexagonal
HoMnO$_3$.$\cite{18}$ Dielectric anomalies associated with spin
reorientation transitions have been observed in several
manganites$\cite{7,19}$ and the significance of the spin-lattice
coupling and magnetoelastic effects have been shown.$\cite{11}$ The
coupling between the f- and d-spins forming a 90 degree angle can be
mediated by the antisymmetric Dzyaloshinskii-Moriya interaction
(that is proportional to the vector product of f- and d-moments and
gives rise to weak ferromagnetism) or by a pseudodipolar interaction
arising from the anisotropy of the f-d exchange.$\cite{13}$ In any
case, the exchange interaction maximizes if the two moments are
perpendicular to each other.

The Fe spin reorientation observed in GdFe$_3$(BO$_3$)$_4$ at
T$_{SR}$ is triggered by this f-d exchange mechanism. We propose
that the observed enhancement of the dielectric constant below T$_N$
(Figs. 1 to 3) is closely related to the indirect f-d exchange
coupling and the magnetic polarization of the Gd moments. The
increase of $\epsilon_a$(T) as T approaches T$_{SR}$ at H=0 reflects
the increased coupling strength and AFM Gd sublattice polarization.
The spin-phonon coupling causes the lattice to soften as the
magnetic order becomes unstable for T $\rightarrow$ T$_{SR}$.  The
combined effects on $\epsilon_a$ saturate at T$_{SR}$(H=0) and it
disappears abruptly with the spin reorientation of the iron at lower
T resulting in the sudden decrease of the dielectric constant. With
the magnetic field applied along the $c$-axis T$_{SR}$ is lowered
but the polarizing effect on the Gd moments and the enhancement of
$\epsilon_a$(T) still reaches its maximum at 9 K explaining the peak
of $\epsilon_a$ remaining H independent at T$_M$$\approx$9 K. With
the magnetic field applied along the $a$-axis the Gd moments are
tilted away from the $c$-axis and the in-plane AFM order of the
Fe-spins is gradually reduced, as can be concluded from the field
effect on the susceptibility, $\chi_a$ (Fig. 4). Both effects reduce
the f-d exchange coupling and the magnetic polarization of the
Gd-spins and its effect on the lattice is reduced. The saturation of
the lattice softness happens at higher temperature in the in-plane
magnetic field, the maximum of $\epsilon_a$(T) is suppressed, and
T$_M$ increases with H (Fig. 3). Although this discussion is
qualitative it does explain the two anomalies observed in the
temperature dependence of the dielectric constant and their opposite
dependence on the magnetic fields parallel or perpendicular to the
$c$-axis. The softening of the lattice at T$_M$ eventually results
in a ferroelectric displacement as observed in recent polarization
measurements$\cite{21}$ that indicate the existence of a sizable
field-induced polarization between T$_{SR}$ and T$_M$ for both
orientations ($c$-axis and in-plane) of the magnetic field. The
corresponding magnetostriction data provide the experimental prove
for the lattice anomalies at the phase boundaries of the FIP phase,
T$_{SR}$ and T$_M$.

The thermodynamic signature of the magnetic phase transitions at
T$_N$ and T$_{SR}$ is given by sharp peaks of the zero-field heat
capacity$\cite{20}$ at the phase transitions. In external magnetic
fields, H $\parallel$ $c$, the heat capacity peak at T$_{SR}$ shifts
to lower T , in accordance with the magnetic and dielectric data
discussed above. Fig. 5 shows the heat capacity close to T$_{SR}$ at
different values of H$_c$. All peaks are relatively sharp and
symmetrical indicative of a first order phase transition. The heat
capacity peak does not shift with H $\parallel$ $a$ which is
consistent with our dielectric and magnetic data. From the
structural point of view the exchange interaction between gadolinium
ions should be weak since they are interconnected by BO$_3$
triangles that are 2.4 ${\AA}$ apart. Thus T$_M$ is not observed in
the heat capacity or magnetization data but only in dielectric data.

The magnetic phase diagram of GdFe$_3$(BO$_3$)$_4$ at lower
temperatures derived from the heat capacity, magnetization and
dielectric measurements is shown in Fig. 6 for both orientations of
the magnetic field. T$_{SR}$ is not affected by H$_a$, however, it
decreases quadratically with H$_c$, in agreement with the results of
recent resonance$\cite{6}$ and magnetoelectric
experiments.$\cite{21}$ The field dependence of T$_M$ is also shown
in Fig. 6 by dashed lines. T$_M$ merges with T$_N$ for
H$_{a}\approx$ 5 kOe in good agreement with the anomalies detected
in polarization and magnetostriction measurements. However, for H
$\parallel$ c, T$_M$(H) is almost constant and extends to far higher
fields than T$_{SR}$. T$_{SR}$ and T$_M$ form the phase boundaries
of the FIP phase where magnetic field-induced polarization was
observed.$\cite{21}$ It is interesting to note that the heat
capacity does not show any anomaly at T$_M$ as it does at T$_{SR}$.
The transition across T$_M$ is obviously not accompanied by
anomalies such as a sizable change of volume etc. This indicates a
more subtle change in the correlated magnetic systems of Gd and Fe
spins coupled to the lattice that results in the distinct peak of
the dielectric constant.

Given the sharp nature of the heat capacity peak anomaly at T$_{SR}$
as well as its sudden drop in magnetization and dielectric constant,
a first order phase transition is presumed to occur. To further
verify the validity of the first order nature at T$_{SR}$ we make
use of the magnetic analog of the Clausius-Clapeyron equation,

$-$$\Delta$S(B)/$\Delta$M(B) = dB/dT$_{SR}$(B)

where $\Delta$S is the change in entropy, $\Delta$M is the change in
magnetization and dB/dT$_{SR}$ is the inverse slope of T$_{SR}$(B)
with respect to the internal magnetic field B. The set of values for
$\Delta$S are obtained from integrating the area underneath the peak
anomaly of Cp/T at different fields (Fig. 5). The set of values for
$\Delta$M is obtained from the change of the magnetization at
T$_{SR}$ (Fig. 4). dB/dT$_{SR}$ is extracted from the phase diagram
of Fig. 6 and B=$\mu$$_0$(H+M). For the case of H $\parallel$ $c$,
equation (1) is well fulfilled as shown in Fig. 7. This proves the
first order nature of the spin reorientation transition.

In conclusion, we have demonstrated a strong correlation between the
magnetic order and the dielectric properties of
GdFe$_3$(BO$_3$)$_4$. Below the AFM ordering temperature of the
Fe-spins, T$_N$, $\epsilon_a$(T) increases and reaches a maximum at
the spin reorientation transition temperature, T$_{SR}$, in zero
magnetic field. We interpret the apparent lattice softness as a
consequence of the indirect exchange coupling between the Gd- and
Fe-spins. For non-zero magnetic fields the $\epsilon_a$(T)-maximum
(at T$_M$) and the spin reorientation transition are separated and
appear at different temperatures. The spin reorientation transition
is proven to be a first order phase transition through the magnetic
analogue of the Clausius-Clapeyron equation.

\begin{acknowledgments}
This work is supported in part by NSF Grant No. DMR-9804325, the
T.L.L. Temple Foundation, the J. J. and R. Moores Endowment, and the
State of Texas through the TCSUH and at LBNL by the DoE.
\end{acknowledgments}


\begin{figure}
\caption{(Color Online) Dielectric data showing all three major
anomalies. T$_1$=156 K, T$_N$=36 K and T$_{SR}$=9 K. Inset shows the
behavior of $\epsilon$(T) around T$_N$ when $\epsilon$ is measured
along the hexagonal $a$- and $c$-axis for H=0.}

\caption{(Color Online) Splitting of the spin reorientation phase
transition under fields parallel to the $c$-axis. T$_{SR}$ shifts
lower while T$_M$ stays the same.}

\caption{(Color Online) Splitting of the spin reorientation phase
transition under fields parallel to the $a$-axis. T$_{SR}$ remains
constant while T$_M$ shifts higher.}

\caption{(Color Online) Mass susceptibility data at different
magnetic fields parallel and perpendicular to the $c$-axis. T$_{SR}$
shifts lower and $\Delta$M increases under magnetic fields parallel
to the $c$-axis (open). T$_{SR}$ stays the same if magnetic fields
are aligned with the $a$-axis (solid). $\chi_a$ is only affected
between T$_{SR}$$<$T$<$T$_N$.}

\caption{(Color Online) Heat capacity data at different magnetic
fields parallel to $c$-axis. T$_{SR}$ shifts lower and $\Delta$S
decreases with H$_c$.}

\caption{(Color Online) Phase diagram of GdFe$_3$(BO$_3$)$_4$ at low
temperatures including T$_{SR}$, T$_M$ and T$_N$ derived from
dielectric data. AFM1 is the state where the Fe spins are ordered
antiferromagnetically along the basal plane, AFM2 is the state where
the Fe spins are ordered antiferromagnetically along the $c$
direction. FIP is the state where there is field induced
polarization.}

\caption{(Color Online) The Clausius-Clapeyron equation satisfied at
different magnetic fields H$_c$.}
\end{figure}

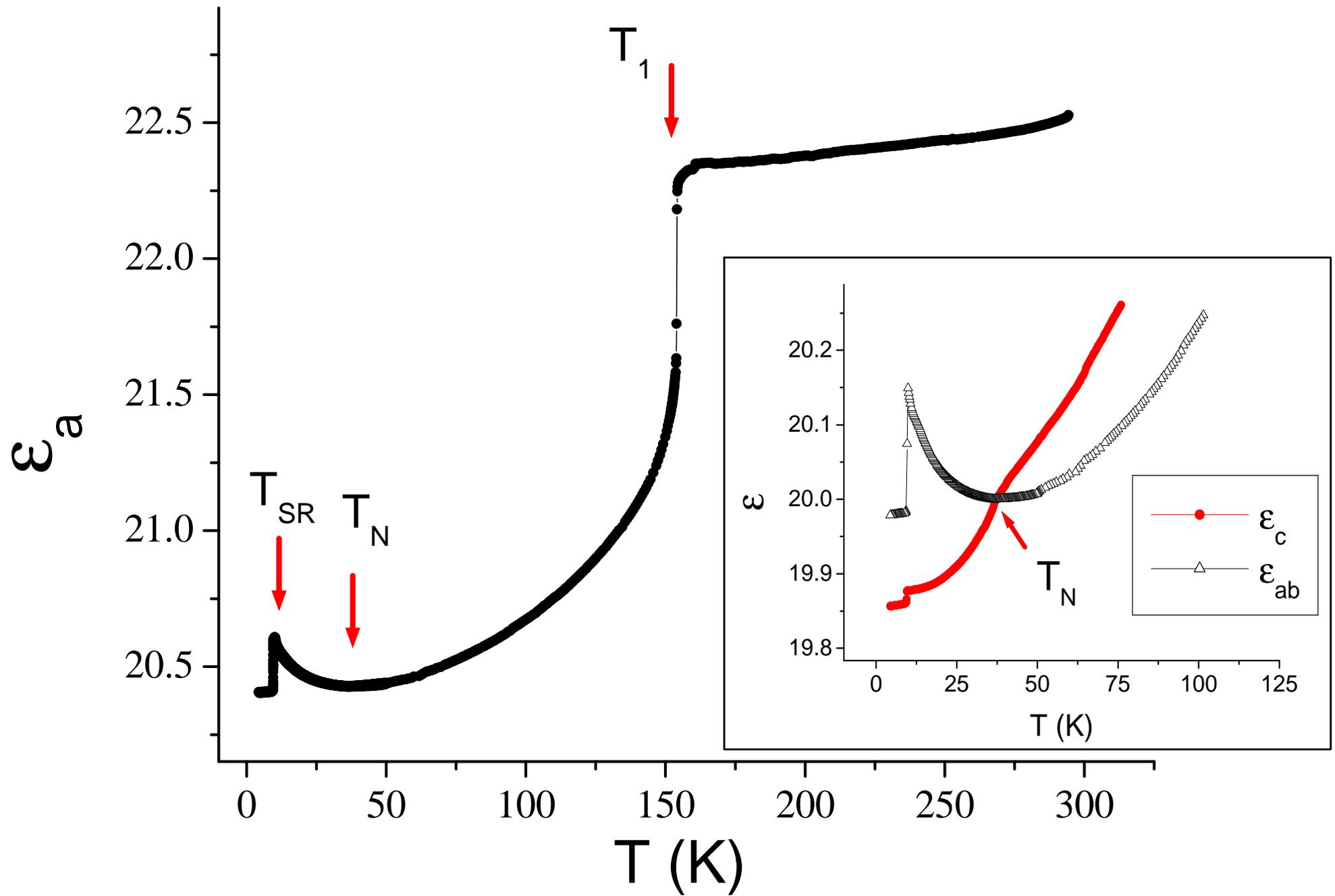
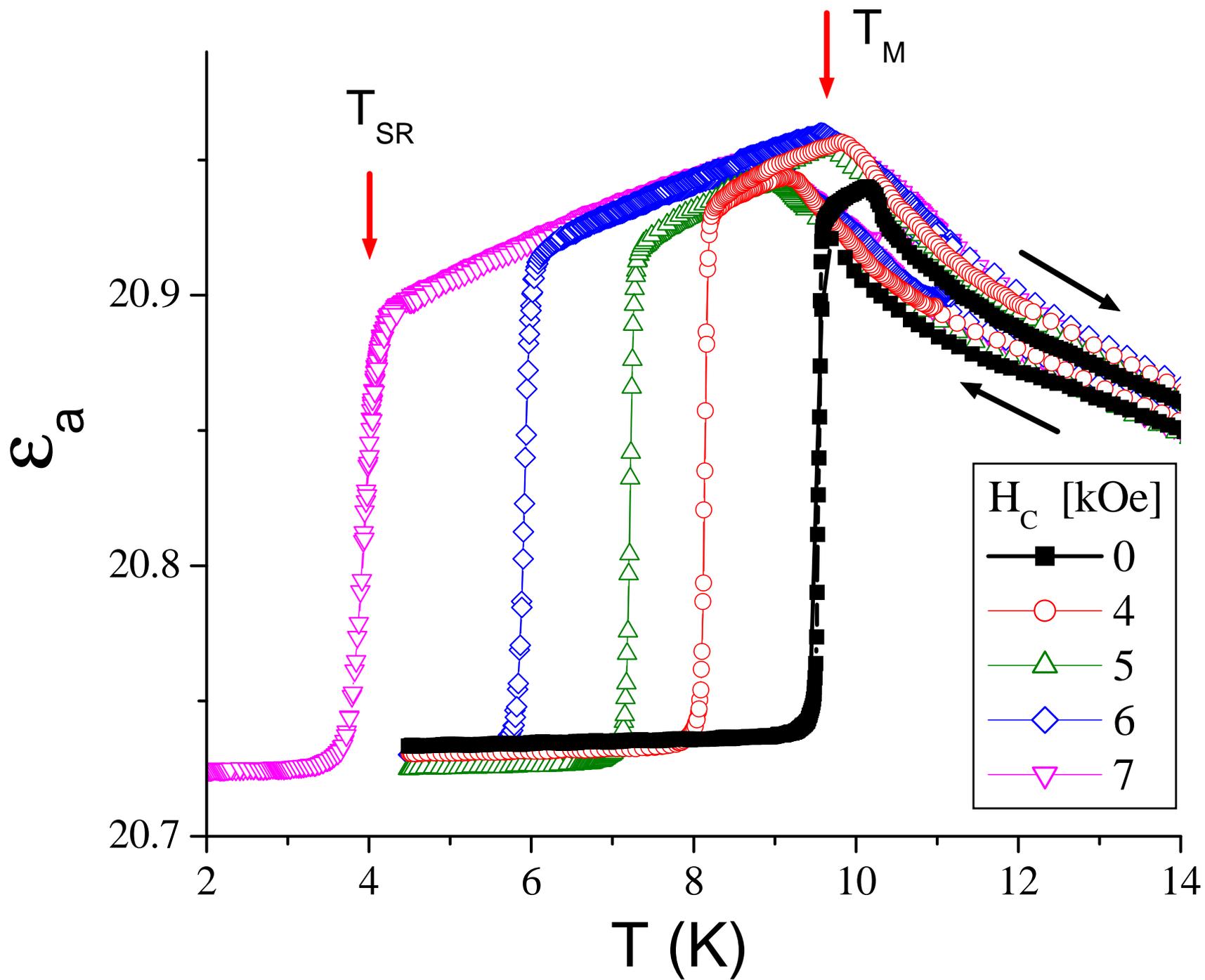
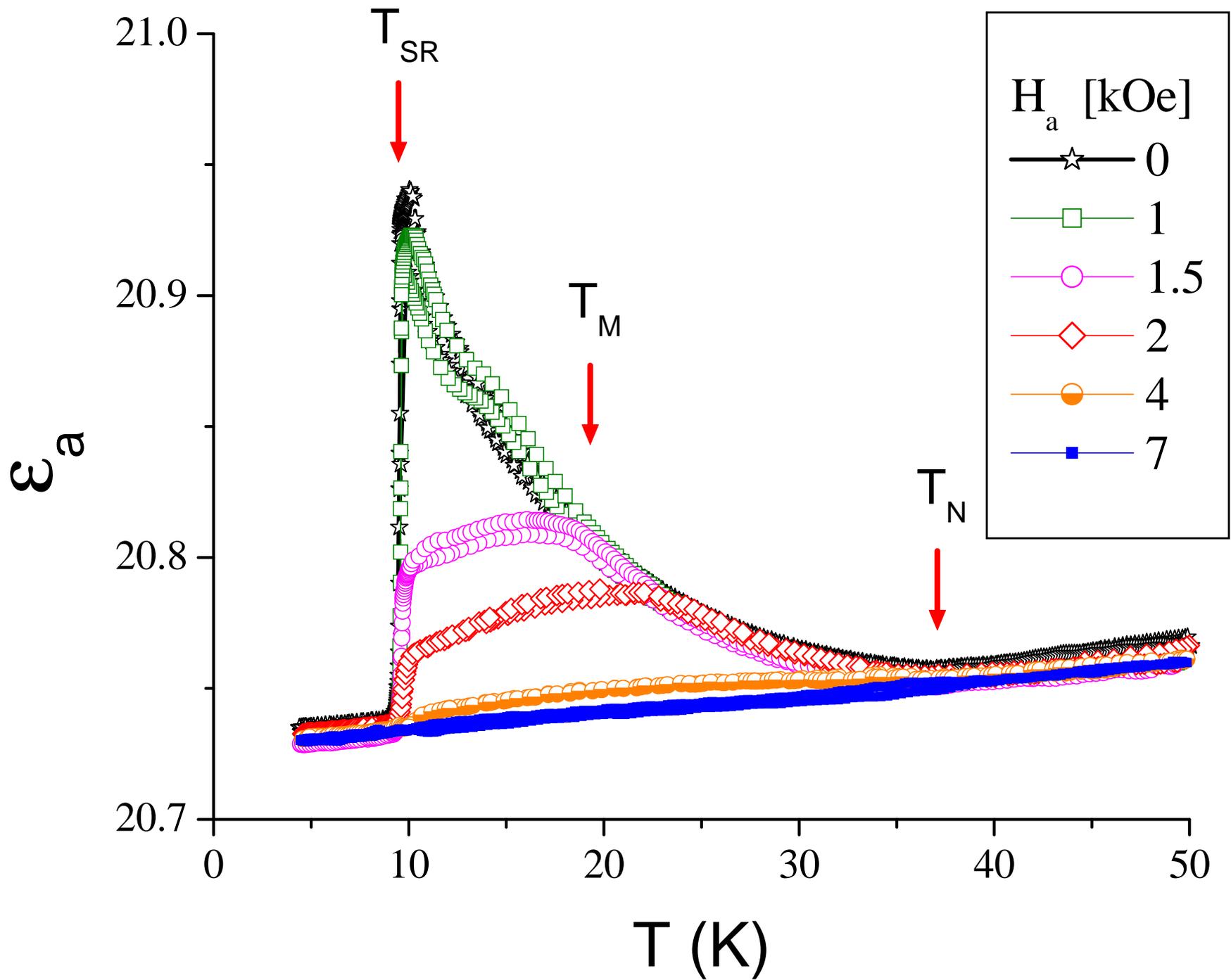
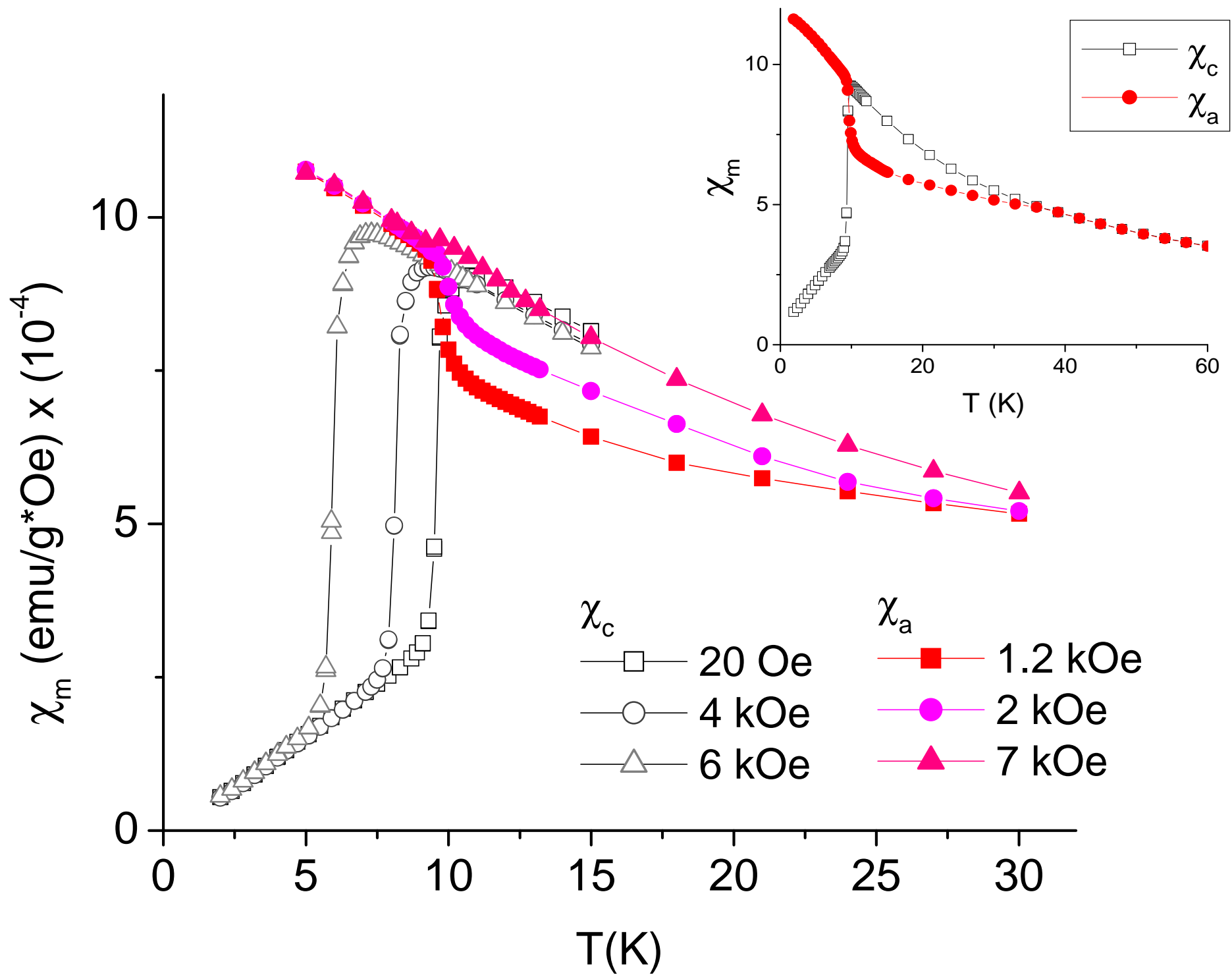
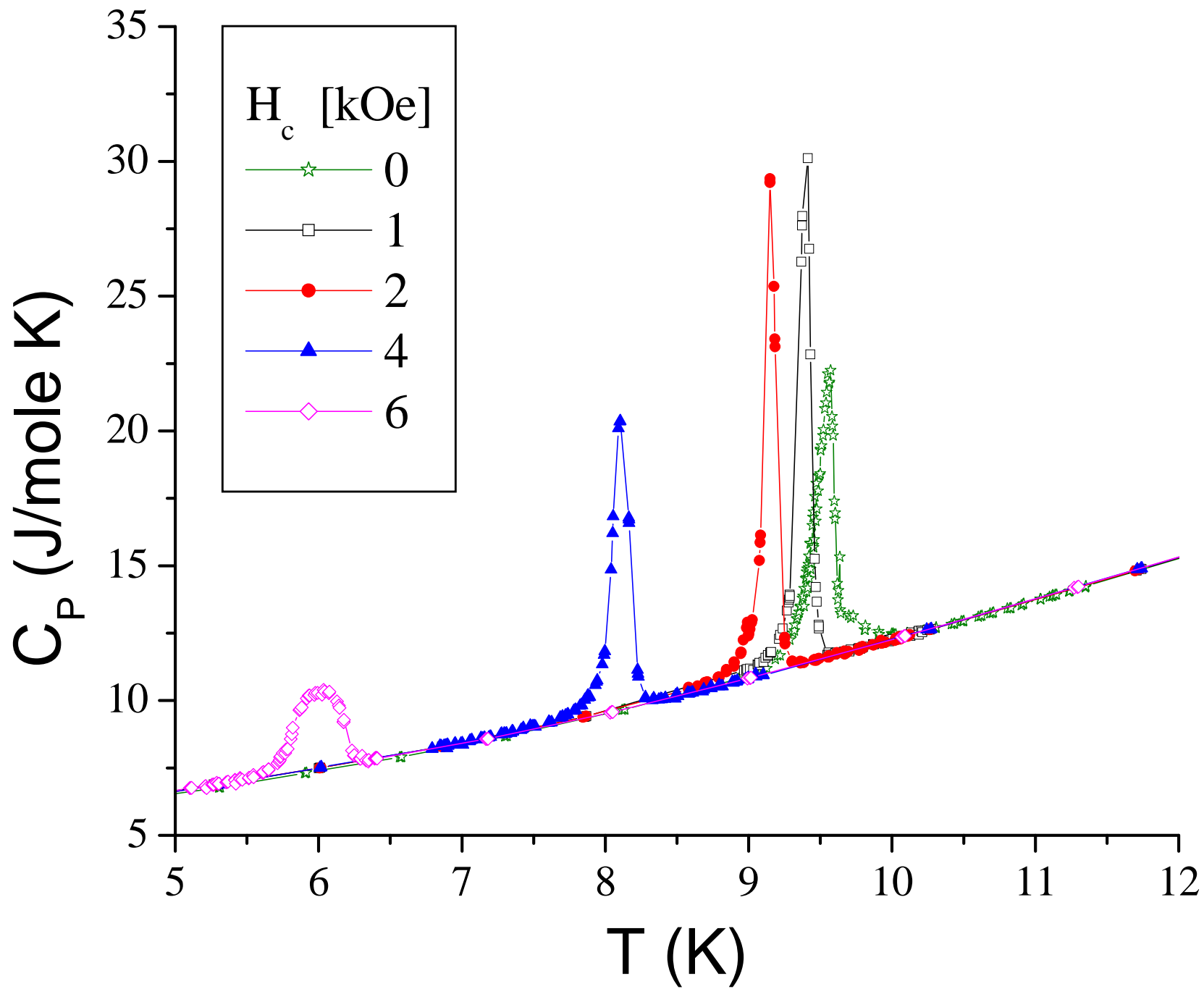
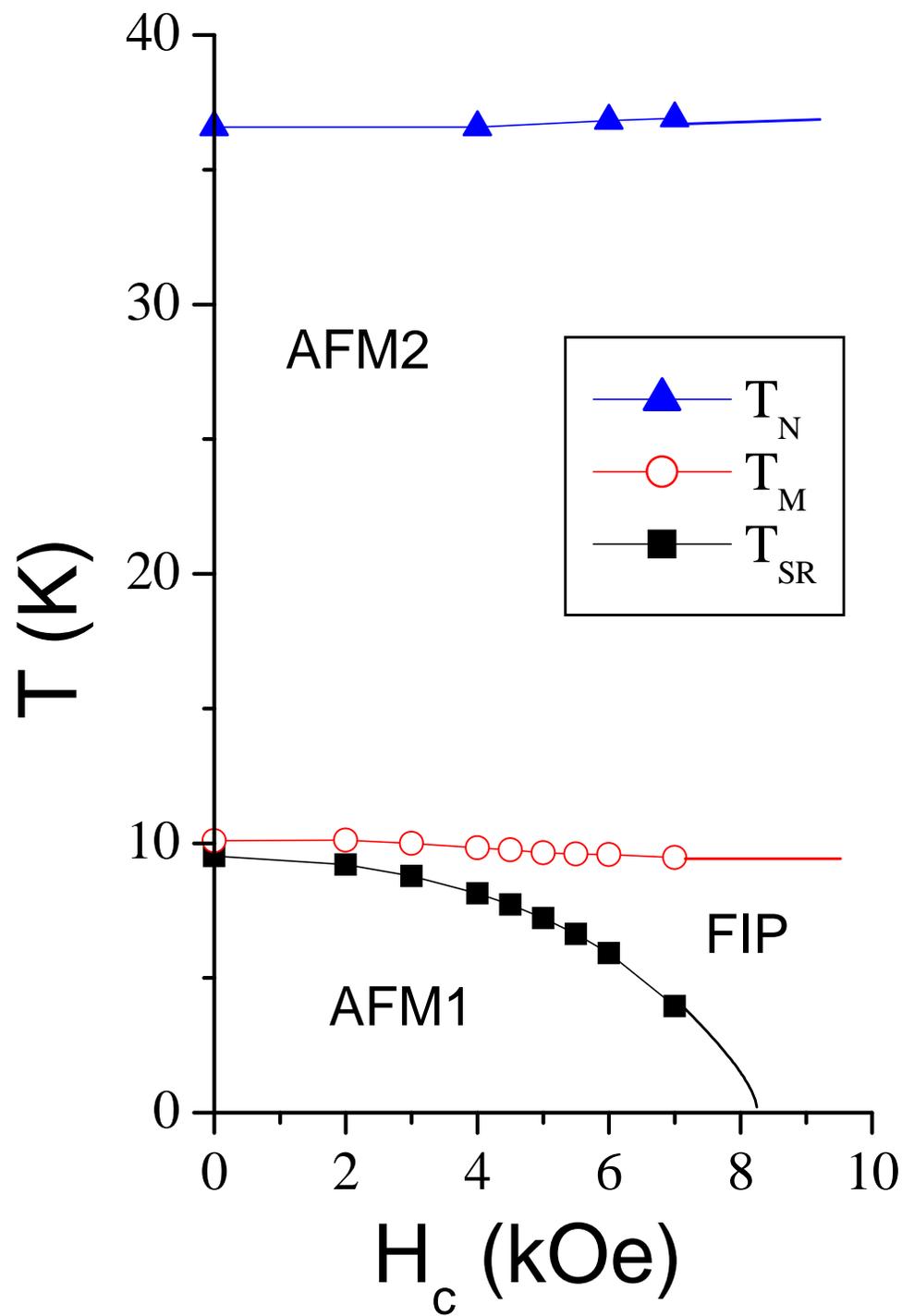
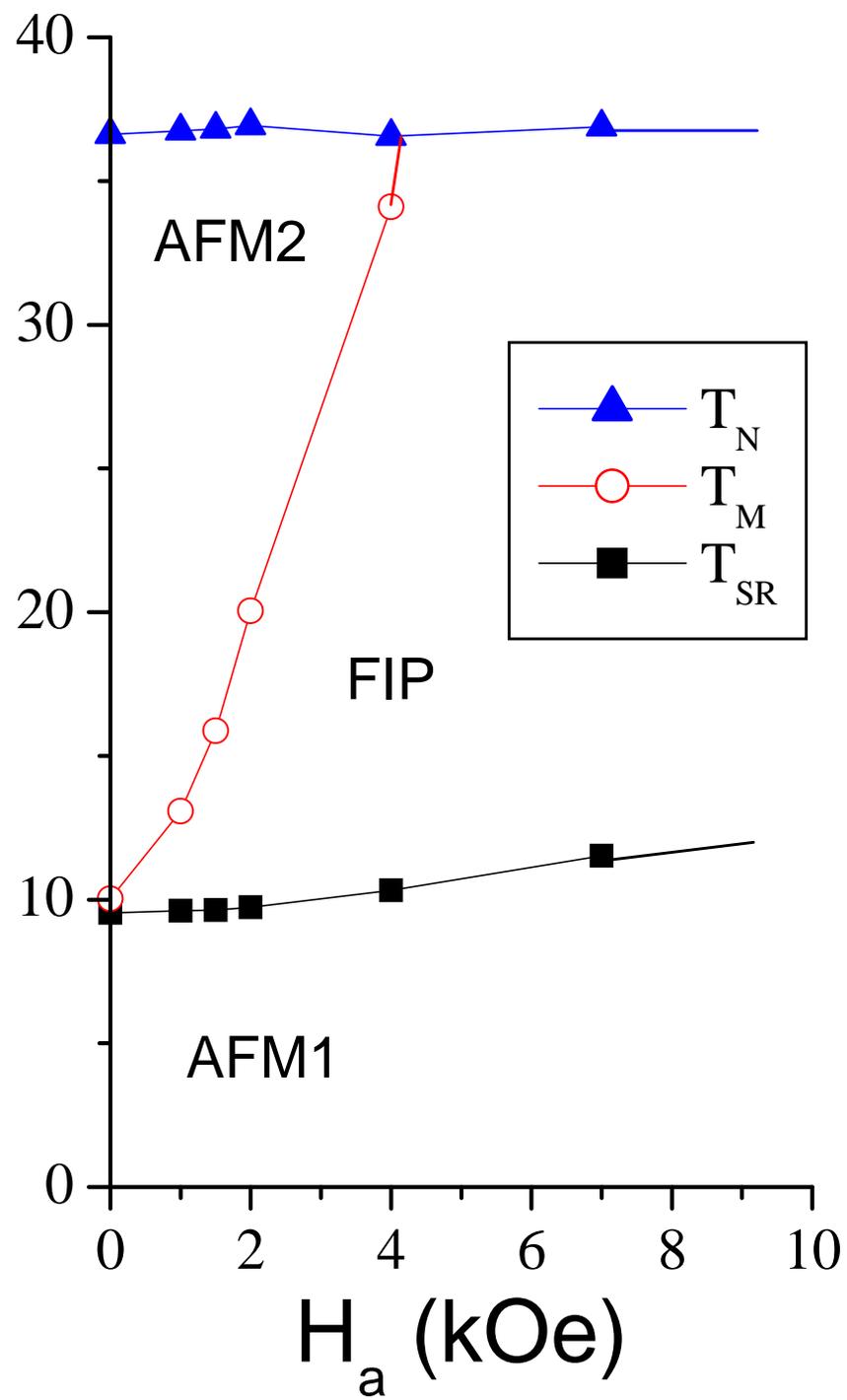
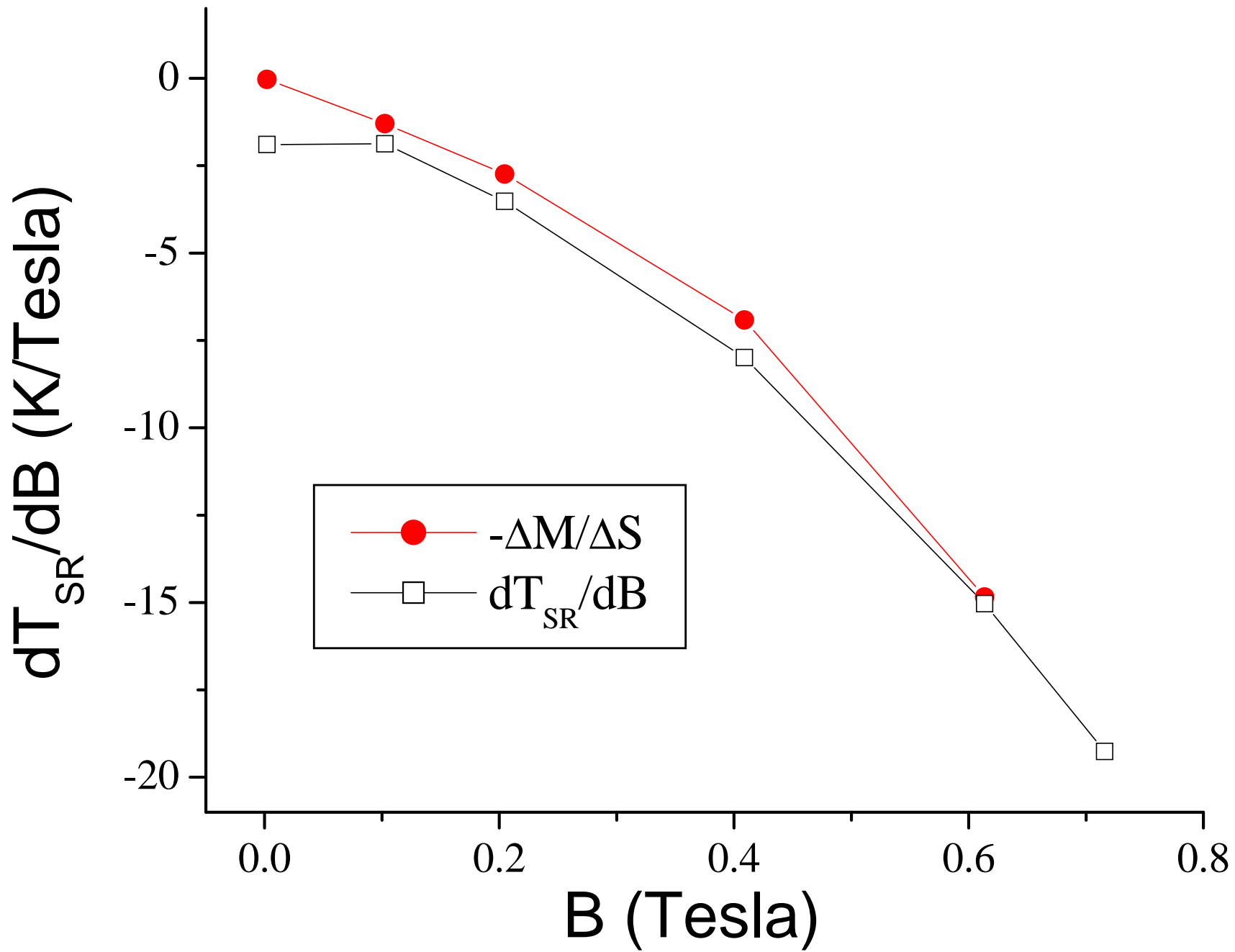
\end{document}